\documentclass[aps,twocolumn,superscriptaddress]{revtex4-1}
\setcitestyle{super}

\usepackage{filecontents}
\usepackage{graphicx}
\usepackage{color}
\usepackage{multirow}
\usepackage{rotating}
\usepackage{hyperref}

\usepackage{pdfpages}
\makeatletter
\AtBeginDocument{\let\LS@rot\@undefined}
\makeatother

\begin{document}

\title{Many-body quantum Monte Carlo study of 2D materials:\\
cohesion and band gap in single-layer phosphorene 
}

\author{T. Frank}
\affiliation{
University Regensburg, Institute for Theoretical Physics, 
93040 Regensburg, Germany
}

\author{R. Derian}
\affiliation{
Center for Computational Materials Science, Institute of Physics, Slovak Academy of Sciences, 84511 Bratislava, Slovakia
}

\author{K. Tok{\' a}r}
\affiliation{
Center for Computational Materials Science, Institute of Physics, Slovak Academy of Sciences, 84511 Bratislava, Slovakia
}

\author{L. Mitas}
\affiliation{
Department of Physics, North Carolina State University, Raleigh, NC 27695-8202
} 

\author{J. Fabian}
\email{jaroslav.fabian@ur.de}
\affiliation{
University Regensburg, Institute for Theoretical Physics, 
93040 Regensburg, Germany
}

\author{I. {\v S}tich}
\email{ivan.stich@savba.sk}
\affiliation{
Center for Computational Materials Science, Institute of Physics, Slovak Academy of Sciences, 84511 Bratislava, Slovakia
}
\affiliation{
Institute of Informatics, Slovak Academy of Sciences, 845 07 Bratislava, Slovakia
}
\affiliation{
Department of Natural Sciences, University of Ss. Cyril and Methodius, 917 01 Trnava, Slovakia.
}

\begin{abstract}
Quantum Monte Carlo (QMC) is applied to obtain the fundamental (quasiparticle) electronic band gap, $\Delta_f$, of a semiconducting two-dimensional (2D) phosphorene
whose optical and electronic properties fill the void between graphene and 2D transition metal dichalcogenides.  Similarly to other 2D materials, the electronic structure
of phosphorene is strongly influenced by reduced screening, 
making it challenging to obtain reliable predictions by
single-particle density functional methods. Advanced
GW techniques, which include many-body effects as perturbative corrections, are hardly consistent with each other, predicting the band gap of phosphorene with a spread of almost 1 eV, from 1.6 to 2.4 eV. Our QMC results, from infinite periodic superlattices
as well as from finite clusters, predict $\Delta_f$ to be about 2.4 eV, indicating that available GW results are systematically underestimating the gap. Using the recently uncovered universal scaling between the exciton binding energy and $\Delta_f$, 
we predict the optical gap of 1.75 eV that
can be directly related to measurements
even on encapsulated samples due to
its robustness against dielectric environment.
The QMC gaps are indeed consistent with recent
experiments based on optical absorption and photoluminescence excitation spectroscopy.
We also predict the cohesion of phosphorene to be only slightly 
smaller than that of the bulk crystal. Our investigations
not only benchmark GW methods and experiments, but also open the field of 2D electronic structure to computationally intensive but highly predictive QMC methods which include many-body effects such as electronic correlations and van der Waals interactions explicitly. 
\end{abstract}


\maketitle

\section{\label{sec:intro}Introduction}

Two-dimensional (2D) materials have already revolutionized science, and have the potential to revolutionize technology due to their
unique electronic, optical, thermal, spin, and magnetic properties.~\cite{li_14,liu_14,novoselov_04,hunt_13,heinz_10,hisao_04,han_14} Remarkably,  2D materials cover a wide range of electronic
structures.
The electronic properties range from metallic single atom layers of palladium and rhodium,~\cite{mao_14} semimetallic graphene,~\cite{novoselov_04} semiconducting transition metal dichalcogenides,~\cite{heinz_10} to insulating wide-gap h-BN.~\cite{hisao_04} Crucial for device applications are materials with a proper band gap. Layered
black phosphorus (BP) features fundamental band gaps in the range of 0.3--2 eV, 
bridging semimetallic graphene and transition metal dichalcogenides. \cite{ling_15} This range 
is specifically important for optoelectronic, photovoltaic, photocatalytic, fiber optic telecommunications, and thermal imaging applications.~\cite{gomez_15}

\begin{figure}
\centering
\includegraphics[width=1.0\columnwidth,clip,angle=0]{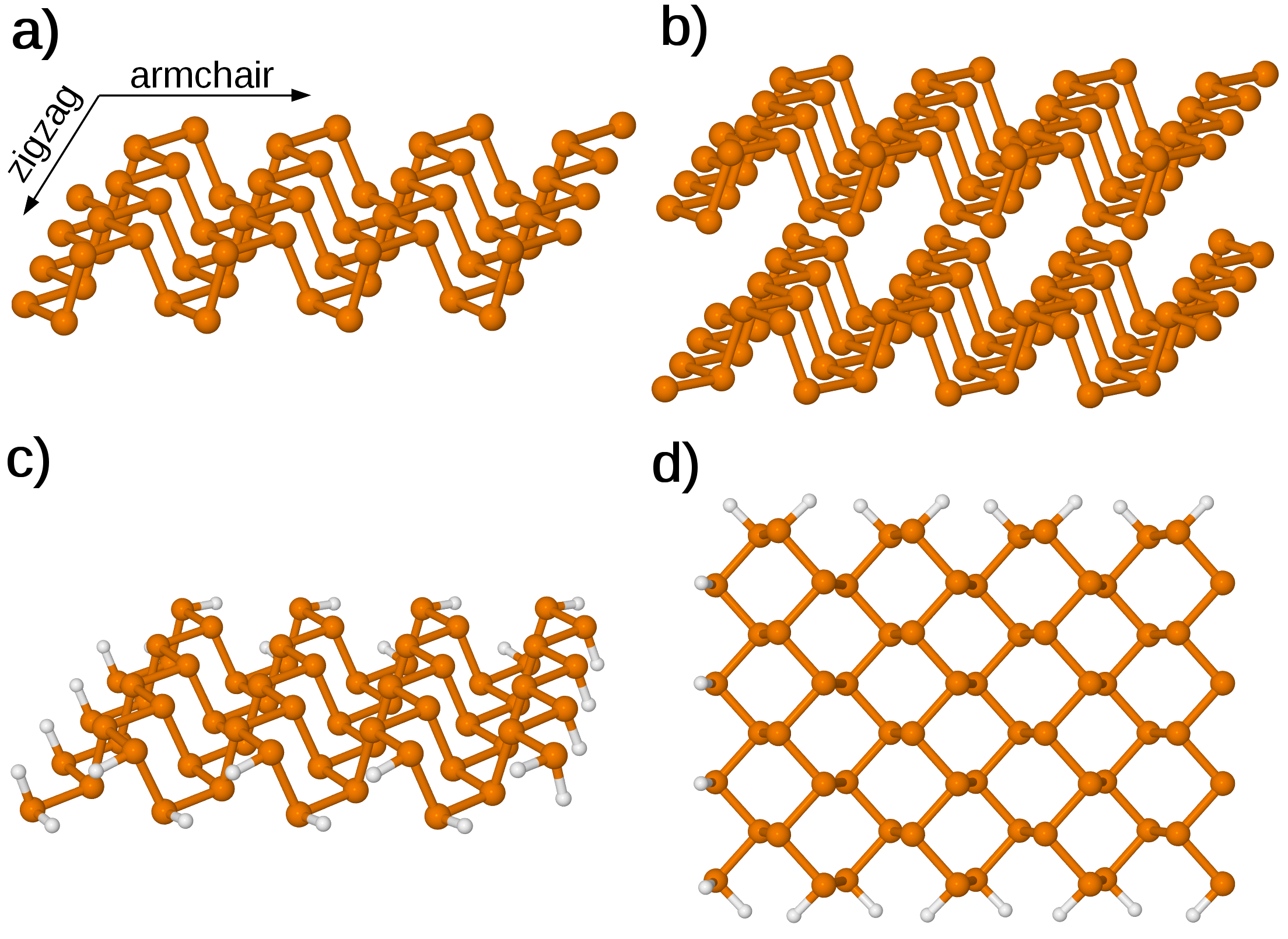}
\caption{
Atomic structure of a) single-layer and b)  few-layer phosphorene. Characteristic armchair and zigzag directions
are indicated in a). Side and top views of a 4 $\times$ 4 cluster approximant, with saturated edge bonds, are presented in c) and d), respectively.
}
\label{fig:struct_mod}
\end{figure}

A single layer of black phosphorus---phosphorene---comprises
sp$^3$ bonded phosphorus atoms forming an anisotropic puckered honeycomb lattice, see Figure~\ref{fig:struct_mod}. The three-fold bonding coordination implies that each phosphorus atom has a lone pair orbital which makes phosphorene reactive to air. \cite{ling_15} This oxidation degradation is eliminated by capping or encapsulating phosphorene with(in) an insulator.   

\begin{table*}
\centering
\caption{
Comparison of selected experimental and computed fundamental ($\Delta_{f}$) and optical ($\Delta_{o}$) gaps and exciton binding energies, in [eV], in unstrained single-layer phosphorene. STS, PL, PLES, and OA stand for scanning tunneling spectroscopy, photoluminiscence, photoluminescence excitation spectroscopy, and optical absorption, respectively. The experimental samples correspond to exfoliation onto different substrates (Si, SiO$_2$, sapphire) and were studied either freshly prepared or capped by h-BN layer. 
$G_{0}$ and $W_{0}$ imply that the Green's function and screened Coulomb repulsion in the GW approach are calculated non-self-consistently. In GW calculations, the exciton binding energies, where quoted, were calculated in the GW-BSE (Bethe-Salpeter equation) approach.~\cite{gatti_16} GGA means generalized gradient approximation, H hybrid DFT functional. The present results appear in bold: DFT with GGA (PBE~\cite{pbe}) and H functional (B3LYP~\cite{B3LYP}) and DMC (upper two entries correspond to the periodic system with B3LYP and PBE nodal hypersurfaces, respectively, while the bottom entry to the cluster system with B3LYP treatment). 
The curly brace indicates compound value due to the cluster upper limit property, see the text.
The star at the exciton binding energy means estimation as $\Delta_{b} = 0.27\Delta_{f}$.~\cite{duan_17}
}
{\small 
\begin{tabular}{c c|c c c c c c c c| c c| c c| c c}
\hline
\hline \multicolumn{14}{c|}{$\Delta_{f}$}&\multicolumn{2}{c}{$\Delta_{o}$}\\
\hline     
\multirow{2}{*} {}DFT & DFT & $GW_{0}$ & $GW_{0}$ & $GW_{0}$ & $GW_{0}$ & $G_{0}W_{0}$ & $G_{0}W_{0}$ & $G_{0}W_{0}$ & $G_0 W_0$ & \multicolumn{2}{c|}{DMC} & exp. & exp. & exp. & exp. \\
GGA & H & GGA & H & GGA & GGA & GGA & GGA & GGA & GGA &\multicolumn{2}{c|}{GGA/H} & STS & PLES & PL & OA \\  
\hline
\multirow{3}{*}{0.8\cite{zhang_14}} & 1.5\cite{zhang_14} & \multirow{3}{*}{1.9\cite{liang_14}} & \multirow{3}{*}{2.4\cite{geng_15}} & \multirow{3}{*}{2.3\cite{choi_15}} & \multirow{3}{*}{2.3\cite{thygesen_16}} & \multirow{3}{*}{2.0\cite{yang_14,thygesen_16}} & 
\multirow{3}{*}{1.8\cite{duan_17}} & \multirow{3}{*}{1.6\cite{rudenko_14}} & \multirow{3}{*}{2.1\cite{ferreira_17}} &  \textbf{2.68$\pm$0.10} &\multirow{3}{*}{$\Bigg\}$\textbf{2.4}} & \multirow{3}{*}{2.0\cite{liang_14}} & \multirow{3}{*}{2.2$\pm$0.1\cite{xia_15}} & 2.1$\pm$0.02\cite{surrente_16}, 1.6\cite{yan_16} &
\multirow{3}{*}{1.7\cite{wang_17}}\\
 &  &  &  &  &  &  &  & & &\textbf{2.54$\pm$0.12} &  &  &  &2.0$\pm$0.04\cite{surrente_16}, 1.5\cite{liu_14}   &  \\  
 \textbf{0.8~~} & \textbf{1.7~~} &  &  & &  &  &  & &  &\textbf{2.41$\pm$0.17} &  &  & &1.3$\pm$0.02\cite{xia_15}&  \\ 
\hline
\multicolumn{16}{c}{$\Delta_{b}$} \\
\hline
 - & - & - & - & 0.9\cite{choi_15} & - & 0.8\cite{yang_14} & 0.5\cite{duan_17} & - &  0.8\cite{ferreira_17} \ &\multicolumn{2}{c|}{\textbf{$\approx $ 0.65}$^{\star}$}& - & 0.9$\pm$0.12\cite{xia_15} & - & - \\  
\hline
\hline
\end{tabular}
\label{tab:previous_gaps} 
}
\end{table*}

Black phosphorus features a direct band gap at the $\Gamma$ point, all the way from single-layer phosphorene up to bulk. 
However, and this makes phosphorene particularly attractive, there are several ways to manipulate the gap. (i) First,
the gap varies with the number of layers and therefore band engineering techniques can be readily applied. Exfoliation of layers from
the bulk BP is straightforward,~\cite{li_14,liu_14,xia_14}
see Fig.~\ref{fig:struct_mod}, due to predominantly van der Waals interlayer bonds.~\cite{shulenburger_15} While 
in the bulk the gap is about 0.3 eV, it increases towards
about 2 eV (we argue for 2.4 eV) in single-layer phosphorene. \cite{gomez_15}  
(ii) Next, phosphorene can sustain large in-plane compressive/tensile strains in excess of about 10\%, compared to just some 2\% in the bulk.~\cite{roldan} This strain engineering
is predicted to affect the band gap by $\approx \pm$50\%.~\cite{stich_16} Finally, (iii) the gap of phosphorene
is predicted to be strongly susceptible to the dielectric environment.~\cite{gomez_15,liu_14, wang_17}

Assuming that the fundamental (quasiparticle)
 band gap $\Delta_f$ depends on the dielectric that protects it against degradation, can we infer from experiments some key characteristics about pristine phosphorene providing thus the benchmark for both experiment and theory? 
Two 
established facts make the answer positive. (a) The dielectric environment affects both the fundamental gap, as well as the exciton binding energy, $\Delta_b$.\cite{zhang_17} Remarkably, the difference, $\Delta_o = \Delta_f - \Delta_b$, which is the optical gap,  is essentially unaffected by the dielectric. \cite{xia_15,thygesen_15,stier_16,veniard_17, raja_17, berkelbach_18} 
(b) Recently, a universal linear scaling 
between the exciton binding energy and the fundamental gap, $\Delta_b \approx 0.27 \Delta_f$ was predicted based on many examples from the 2D realm~\cite{duan_17} (see also the 
predecessor\cite{choi_15}), including phosphorene. Combining these two observations,
(a) and (b), allows to estimate $\Delta_o$ from 
a calculation of $\Delta_f$ on a {\it pristine} 2D material, namely $\Delta_o \approx 0.73 \Delta_f$, and compare with experimental $\Delta_o$ obtained 
from an {\it encapsulated} or {\it capped} sample. 

Let us now turn to the existing experimental 
and theoretical state-of-the-art in determining the electronic gap(s) of single-layer phosphorene. Table~\ref{tab:previous_gaps} shows a rather comprehensive selection of measured and calculated fundamental and optical gaps, as well as exciton binding energies for  phosphorene. Photoluminescence is often 
contaminated by defect emission, as is also evidenced by the scatter of the measured values for $\Delta_o$. Most reliable is optical absorption. A recent experiment\cite{wang_17} has reported $\Delta_o \approx 1.73$ eV, for phosphorene encapsulated in hBN. This value would lead to $\Delta_f \approx 2.4$ eV for pristine phosphorene, according to the above mentioned linear scaling. Similarly, it would suggest the exciton binding energy, $\Delta_b$, of about 0.65 eV. Certainly both (a) and (b) observations are not exact, so the above estimates of $\Delta_f$ and $\Delta_b$ would carry a scatter of perhaps
10\% or so. If we next look at the photoluminescence excitation spectroscopy data 
for phosophorene on silicon oxide, \cite{wang_15} the obtained fundamental 
gap is $\Delta_f \approx 2.2 \pm 0.1$ eV. Considering the influence of the oxide, 
it is reasonable to deduce from this experiment that the fundamental gap of free-standing phosphorene is 0.1--0.2 eV greater, that is, 2.3--2.4 eV. \cite{wang_15}

On the theory side, we see from Tab. I that 
single-particle density functional methods predict, unsurprisingly, too low and strongly 
method-dependent values for $\Delta_f$. Inclusion of GW corrections\cite{gatti_16} is essential to bring the gaps closer to 2 eV.  
However, various GW approximants ($G_{0}W_{0}$ or
$GW_0$) give different values coming from different implementations, ranging from 1.6 eV to 2.4 eV. The largest value, 2.4 eV, which would be consistent with the aforementioned optical absorption experiments, is obtained by using a hybrid functional.\cite{wang_15} However, the same implementation predicts 0.6 eV band gap for bulk BP\cite{wang_15} (experimental value is 0.3 eV), making 
it clear that there is a limited predicting power from this calculation. 
As for the exciton binding energies, predictions (see Tab. I) range from 0.5 to 1 eV,
again with little consensus in both  theory and experiment. The experimental value
of 0.9 eV \cite{wang_15} is likely affected by the optical edge of 1.3 eV of the emission peak (compare to 1.73 eV of the absorption experiment\cite{wang_17}).

To obtain accurate bounds and reliable estimates of the band gap of phosphorene, 
we propose to employ the quantum Monte Carlo (QMC) method. QMC is an efficient, albeit 
computationally demanding ~\cite{QMC_method,needs_10} way to benchmark electronic structure calculations in condensed matter.~\cite{kolorenc_11,wagner_16,dubecky_16}
Indeed, this method has been applied to compute band gaps in three-dimensional systems,\cite{kolorenc_11} clusters,\cite{stich_12} and nanoparticles.\cite{stich_17} In the 2D realm it was already used to obtain reference binding energies of 2D bilayers,~\cite{shulenburger_15,drummond_15} but thus far has not been
systematically employed to obtain electronic structure parameters.

Here we report QMC calculations for the fundamental band gap of single-layer
phosphorene. We use both the periodic lattice as well as supercell approaches, to 
demonstrate convergence and consistency. The accuracy of the ground-state properties is evidenced by calculating cohesion, which differs little from the bulk value. We stress that
our QMC calculation is the full method, not relying on phenomenological interactions. 
In fact, {\it we solve the Schr\"{o}dinger equation for hundreds of electrons} interacting mutually as well as with the lattice ions, to obtain the ground and excited states
needed for the band gap calculation. 
Finally, we note that the knowledge of the band gap of phosphorene is important not only on its own right as a fundamental electronic quantity
of a potentially technologically relevant material, but it is crucial also for
building effective theories such as tight-binding \cite{rudenko_14} and
k$\cdot$p models. \cite{li1_14} We believe that our adaptation of QMC methods will open the way for this powerful technique to investigations of electronic structures of 2D systems, which are inherently prone to strong interactions and require
careful considerations.  

\section{\label{sec:techniques}Simulation techniques}
The band gap $\Delta_{f}$ was determined from both extended and cluster approximants
with lattice parameters fixed to the experimental values in the black phosphorus crystal. In fact, it was shown that the experimental lattice parameters
agree with the lattice parameters determined by QMC methods within the error bars~\cite{shulenburger_15}. The gap
$\Delta_{f}$ was extracted as the singlet-singlet vertical excitation energy. Here $\Delta_{f} \approx E_{v}^{ss} = E_{1}^{s} - E_{0}^{s}$, with $E_{0}$ and $E_{1}$ being, respectively, the ground- and the first excited-states obtained by fixed-node QMC~\cite{QMC_method} not allowing any relaxation of the DFT nodal hypersurfaces due to the HOMO$\rightarrow$LUMO electron excitation; no vibronic effects are included.
The fixed-node approximation is the only fundamental approximation in the electronic structure QMC. \cite{QMC_method}

In the periodic setup the $E_{0}$ and $E_{1}$ were computed from DMC (diffusion Monte Carlo) energies in the fixed-node approximation using the VMC (variational Monte Carlo) trial wave functions with the nodal hypersurfaces determined by two different sets of DFT orbitals: the generalized gradient approximation PBE~\cite{pbe} (PBE/DMC) and hybrid B3LYP~\cite{B3LYP} (B3LYP/DMC), at the $\Gamma$-point of the Brillouin zone, optimizing the short-range correlations of the Jastrow factor.~\cite{QMC_method} 
The consistency check using both PBE and B3LYP DFT nodal hypersurfaces was deemed important as at the DFT level the HOMO-LUMO gaps of the two DFT functionals differ by $\approx$1 eV, see Table~\ref{tab:previous_gaps}. 
The Yeh-Berkowitz~\cite{yeh_99} modification in the 3D Ewald summation technique for systems with a slab geometry that are periodic in two dimensions and have a finite length in the third dimension, was adopted. We cross checked in detail that this agrees with an alternative derivation of Ewald sums for 2D slab geometries.\cite{foulkes_04}
In the cluster setup we used the B3LYP~\cite{B3LYP} (B3LYP/DMC) nodal hypersurfaces.

Finite-size scaling towards the thermodynamic limit was performed for a series of 1 $\times$ 1 to 6 $\times$ 6 series of $L \times L$ periodic approximants,
see Fig.~\ref{fig:struct_mod} {a)}, and for 4 $\times$ 4 and 5 $\times$ 5 H-terminated cluster approximants, assuming a linear scaling with 1/$N$, where $N = 4 \times L \times L$ is the number of P atoms. Our approach corresponds to quasi-exact many-body treatment, to within the fixed-node approximation, of the 2D electron polarizability entering the equations for $\Delta_{f}$.~\cite{duan_17} 
The ground-state energy $E_{0}$ was also used to determine the cohesion energy.  
More details can be found in Supplementary Material (SM).  

\section{\label{sec:results}Results and discussion}

\begin{figure}
\centering
\includegraphics[width=1.0\columnwidth,clip,angle=0]{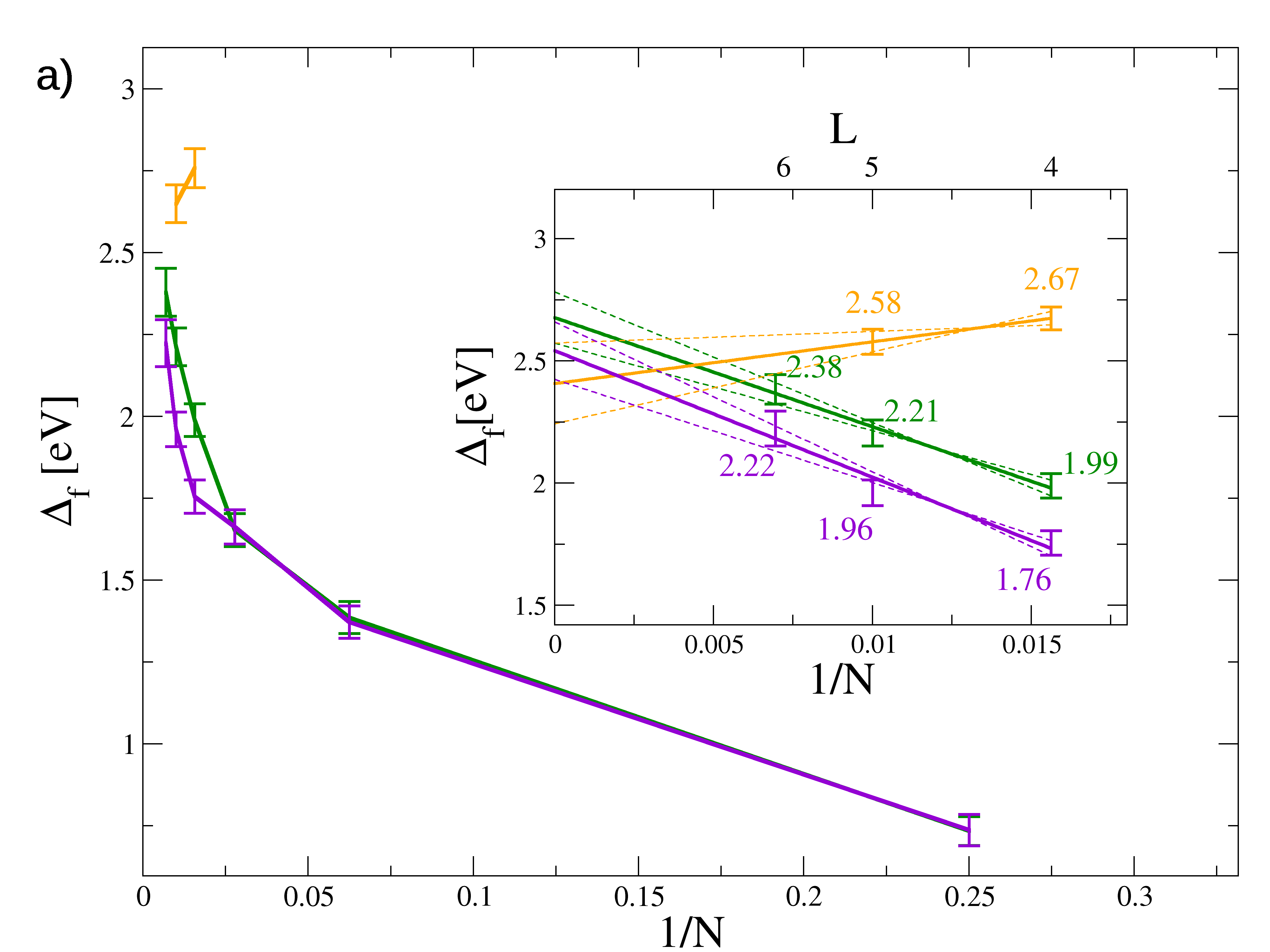}
\includegraphics[width=0.45\columnwidth,clip,angle=0]{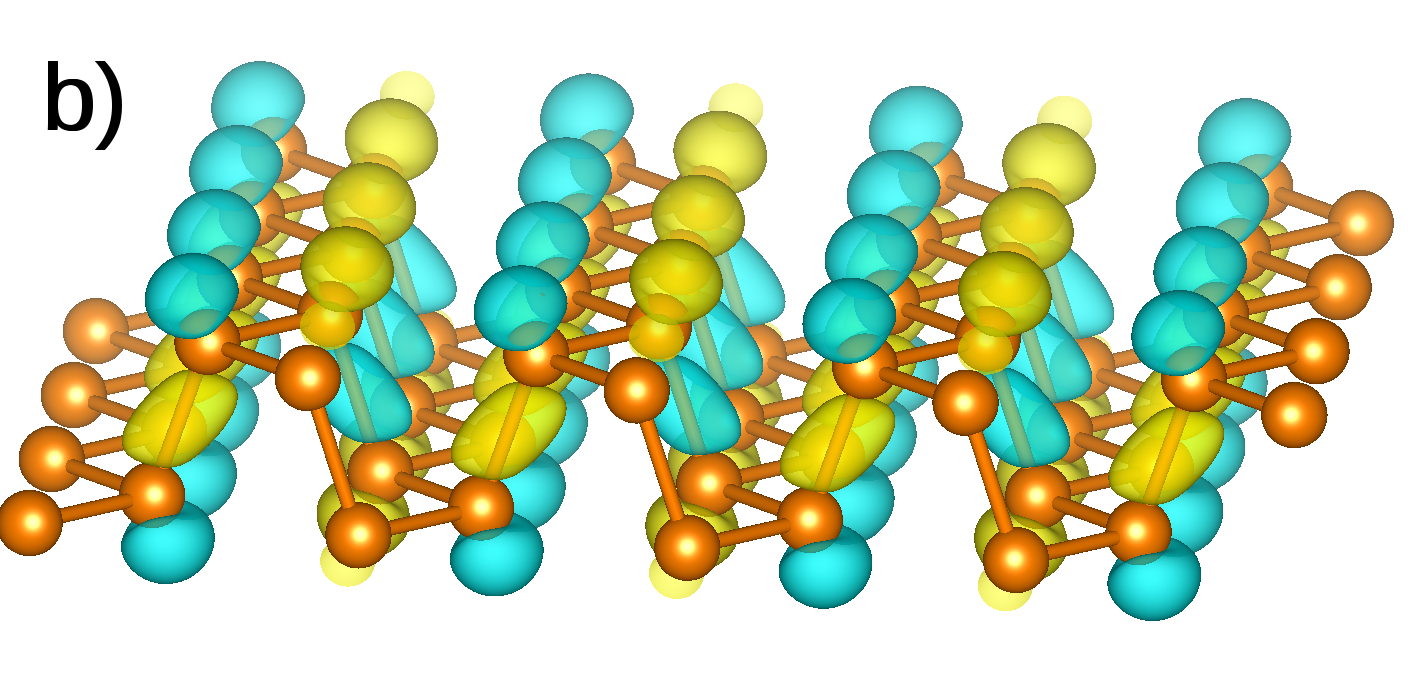}
\includegraphics[width=0.45\columnwidth,clip,angle=0]{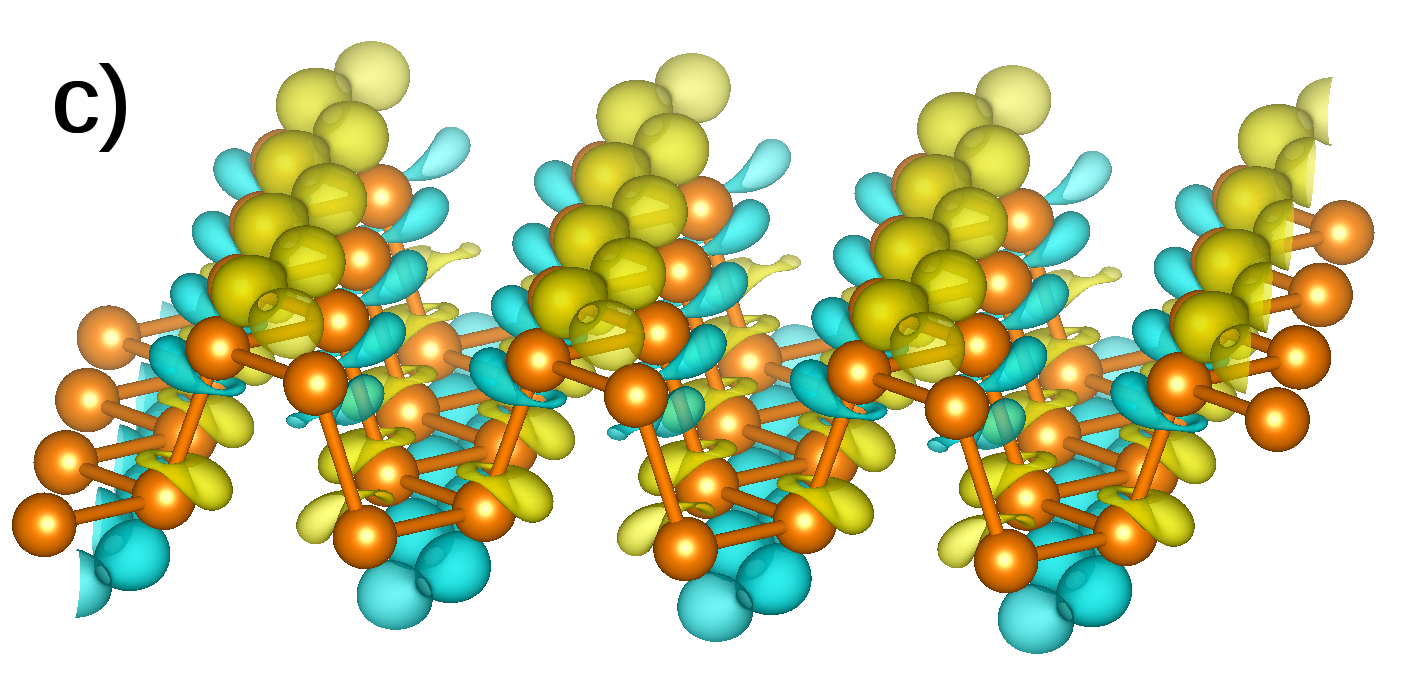}
\caption{
{a)} Finite-size scaling of the computed fundamental gap, $\Delta_{f}$, with the number of atoms, $N$, in the periodic supercell (green line, DMC/B3LYP; purple line, DMC/PBE) and in the cluster approach (yellow line, DMC/B3LYP) along the series of $L \times L$ supercell approximants, $L$ = 1 to 6, for the periodic setup and $L$ = 4, 5 in the cluster approach.
The inset shows the zoom-in of the scaling for large $N$ with the dashed lines showing the linear extrapolation to the infinite size limit.
{b)} B3LYP DFT HOMO (left) and {c)} LUMO (right) orbitals. Both are $\Gamma$-point Bloch states, HOMO being a superposition of bonding orbitals (mostly $\sigma(p_z)$) along the vertical P--P bonds while LUMO being its antibonding counterpart.
}
\label{fig:finite_size_scale}
\end{figure}

\paragraph*{Periodic supercells.}
Our finite-size diffusion Monte-Carlo (DMC) scaling study of $\Delta_{f}$ is shown in Fig.~\ref{fig:finite_size_scale}~{a)}. Although in 3D periodic calculations typical QMC extrapolations from supercells to bulk behave as 1/$N$ where $N$ is the number of atoms in the supercell, in the 2D slab systems the issue is more complicated.~\cite{foulkes_04}
The strong periodic Coulomb interactions unscreened in the direction perpendicular to the layer make the renormalization of electron interactions huge and the electrostatic energies slowly convergent.  
Therefore, the supercells have to be sufficiently large to marginalize the periodicity effects and to allow for reliable finite-size extrapolations.  
To this end, we have carried out extensive calculations of $\Delta_{f}$ for the sequence of 1 $\times$ 1 up to  6 $\times$ 6 periodic supercell systems with two DFT nodal hypersurfaces. For sizes $3 \times 3$ or larger the linear scaling describes the results very well. For smaller supercells, the different components of the total energies, such as the potential energy, are too biased to make the trends transparent.

The infinite-size extrapolation from the extended-system calculations yields 
\begin{equation}
\Delta^{\mathrm{ext},1}_{f} = ~2.68\pm0.10~{\rm eV},
\end{equation}
fixing the nodal hypersurfaces by the hybrid B3LYP~\cite{B3LYP} DFT functional. Recall
that the corresponding DFT value is 1.7 eV, see Tab. I. If we now fix the nodal hypersurfaces
by PBE~\cite{pbe} DFT single-particle orbitals, see Figure~\ref{fig:finite_size_scale} {a)},
we get the extrapolation to 
\begin{equation}
\Delta^{\mathrm{ext},2}_{f} =~2.54\pm0.12~{\rm eV}, 
\end{equation}
which overlaps with $\Delta^{\mathrm{ext},1}_{f}$ within the error bars. The PBE DFT gap is
0.8 eV. Remarkably, despite the significant difference of almost 1 eV in the DFT gaps, the QMC method
that starts with the corresponding DFT wave functions (and fixing their nodal hypersurfaces),
gives a consistent output for both! Surprisingly, the convergence of $\Delta_{f}$ in the periodic setting is determined mainly by the Hartree-Fock energy components, 
with the rest, including the correlation energies, 
converging much faster. For example, the ground-state correlation energy is essentially converged at the 3 $\times$ 3 supercell size, whereas the excited-state at a slightly larger size of 5 $\times$ 5. This provides a way of estimating the phosphorene properties from the more slowly converging Hartree-Fock results, simply by adding the corresponding correlation energies; see SM for more details. We also plot the HOMO/LUMO orbitals for the extended calculations at the $\Gamma$ point in Fig.~\ref{fig:finite_size_scale} b) and c), respectively. The orbital's distribution and bonding type can give insight into Coulomb finite size effects; see SM for a discussion.

\paragraph*{Clusters.} Due to intricacies of extrapolations of the periodic supercells, 
and for consistency, we have also calculated $\Delta_{f}$ in a finite cluster setting. 
The clusters corresponded to supercell sizes in the periodic method above, with terminations of unsaturated bonds with H atoms, see Fig.~\ref{fig:struct_mod} {c)} and {d)}.
Finite-size scaling constructed from these cluster approximants, given in Fig.~\ref{fig:finite_size_scale} {a)}, extrapolates to 
\begin{equation}
\Delta^{\mathrm{clst}}_{f} =~2.41\pm 0.17~eV, 
\end{equation}
reasonably close to $\Delta^{\mathrm{ext}}_{f}$ considering the fact that we compare periodic systems versus isolated clusters in vacuum and that these two models actually exhibit opposite trends as the functions of the system size. Due to finite-size confinement, the computed excitations in finite clusters as a rule overestimate the gaps. The DMC values for the 5 $\times$ 5 clusters clearly illustrate this, see
Fig.~\ref{fig:finite_size_scale}~{a)}. However, as explained in SM, the cluster gaps are bound to converge to the fundamental gap in the infinite size limit and also provide upper bounds for the estimation of fundamental gaps as indicated above.
The upper bound property of cluster gaps enables us to probe usefulness of alternative extrapolations to bulk schemes, such as, for instance, 
1/$\sqrt N$ scaling,~\cite{drummond_13} see SM. 
Since both our estimates of $\Delta_f$ from the periodic supercells are greater than the cluster estimate, there is
likely a small bias in the evaluation of excitations in 
the periodic supercells. In particular, 
the fixed-node errors in the excited and the ground states are most likely not identical and could be also intertwined with the remnants of the finite size errors even at larger sizes.

The convergence of the excited states is much more challenging in extended 2D systems than in bulk. The key reason is that the non-periodic direction enables changes in the single-particle densities that have dominant contribution to the periodic Coulomb interactions, a problem that is naturally avoided in cluster geometries; for details see  SM.
Therefore, we estimate that a systematic uncertainty of 0.1 to 0.2 eV could be present in our periodic-supercells gap calculations. The multiple calculations of the excited states that we present clearly illustrate this aspect. 

The error is certainly greater for the excited-state energy, than for the ground state. Since the variational principle behind QMC 
gives the upper bound on the energy, it follows that the cluster
value is a superior estimate for the fundamental gap. We believe that the true value of $\Delta_f$ for intrinsic phosphorene is
about 2.4 eV, as also indicated in Tab. I. The GW values, defining the current state-of-the-art, are, compared to QMC,
widely scattered and systematically underestimating the gap.

\paragraph*{Exciton binding energy, optical gap, and comparison with experiment.}
Using the universal linear scaling,\cite{duan_17} $\Delta_b \approx 0.27 \Delta_f$, we can also estimate the excitonic
binding energy of phosphorene as $\Delta_b \approx 0.65$ eV. 
The optical gap is then $\Delta_o = \Delta_f - \Delta_b 
\approx 1.75$ eV. This is consistent with the optical absorption 
experiment\cite{wang_17} which reports 1.73 eV. As we discussed
in the introduction, the optical gap of 2D semiconductors should
be insensitive to the dielectric environment, justifying the 
consistency claim. 

In terms of the fundamental gap obtained in experiment, 
(see Tab.~\ref{tab:previous_gaps}), photoluminescence
emission spectroscopy (silicon substrate, no capping),~\cite{xia_15} optical absorption (sapphire substrate, h-BN capping),~\cite{wang_17} and scanning tunneling spectroscopy (no capping),~\cite{liang_14} yield values of $\Delta_{f}$ of 2.2, 1.8, and 2.0 eV, respectively. The optical absorption value was attributed to the h-BN capping which also seems to reduce the exciton binding energy to just 0.1 eV.~\cite{wang_17} The photoluminescence value of 2.2 eV is closest to our QMC prediction. Considering that the sample
was on a dielectric substrate which lowers $\Delta_f$ by perhaps
0.1--0.2 eV,\cite{wang_17} our result is also consistent with this experiment.

\begin{figure}
\centering
\includegraphics[width=1.0\columnwidth,clip,angle=0]{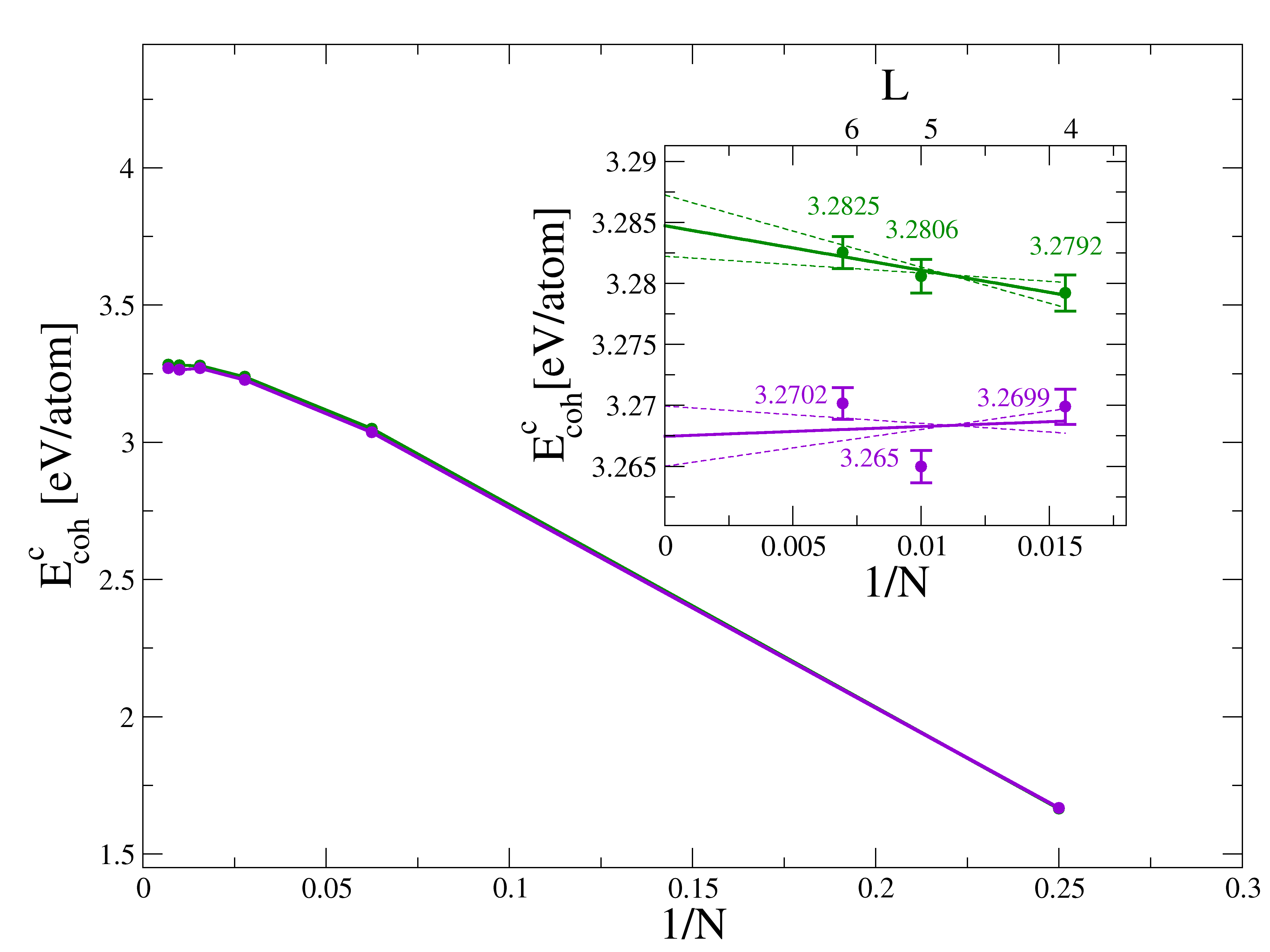}
\caption{
Finite-size scaling of fixed-node corrected cohesion energy
$E_{\rm coh}^c$, in DMC/PBE (purple curve) and DMC/B3LYP (green curve) treatment. The error bars are smaller than the size of the points. The inset shows the zoom-out of the linear scaling for large $N$ with extrapolation to infinite system size.
}
\label{fig:finite_size_scale_cohesion}
\end{figure}

\paragraph*{Cohesion energy.} One important quantity which, to the best of our knowledge, has not been determined experimentally yet, is the free-standing phosphorene cohesion energy. The cohesion energy of bulk black phosphorus is $3.43~\mathrm{eV/atom}$.~\cite{kittel_05} Compared to ordinary semiconducting 3D materials the phosphorus crystal is quite different being a van der Waals system of covalently bonded slabs. Even though the system shows moderately large cohesion, the stability of black phosphorus is rather low mainly due to an easy detachment of $P_4$ molecular units that per atom are bonded almost as strongly as the bulk material. Having calculated the phosphorene DMC ground state energies for a variety of sizes we can estimate the cohesion energy in the thermodynamic limit. Since the P$_{4}$ molecule is the dominant sublimation product from phosphorus solids, to estimate the cohesion we follow the thermodynamic path P$_{4}$ molecule $\rightarrow$ bulk black phosphorus $\rightarrow$ phosphorene using also the QMC calculation of the van der Waals binding energy estimated recently. \cite{shulenburger_15} 
The finite-size scaling, Fig.~\ref{fig:finite_size_scale_cohesion}, yields a fixed-node corrected cohesion energy of 3.268(4) in DMC/PBE and 3.284(6) eV/atom in the DMC/B3LYP periodic calculations,
while from the clusters the estimated value is 3.26(1) eV/atom, see the SM for more details. Note that our estimated DMC values of cohesion energy of phosphorene from the two models are almost identical and very close to the estimation of 3.35 eV/atom for the 2D system obtained by using 
the van der Waals interlayer bonding from another DMC calculation.\cite{shulenburger_15}
The corresponding DFT values are 3.09 eV/atom~\cite{jones_94} and 3.45 eV/atom.~\cite{jeng_16}

\section{\label{sec:conclusion}Conclusion and perspectives}
We have performed systematic fixed-node QMC calculations of the quasiparticle band gap of free-standing single-layer phosphorene in both periodic and cluster settings. Using 
the universal linear scaling between the gap and the optical 
binding energy, we have also extracted the optical gap, 
which can be compared with experiments done on phosphorene on dielectric substrates or on encapsulated samples. Our results
are consistent with available optical absorption and 
photoluminescence emission spectroscopy experiments.
We argue that previous calculations based on GW underestimate
the quasiparticle gap and do not give consistent predictions
for phosphorene. Our ground state is essentially exact, evidenced from the calculated cohesion and its agreement
with available (indirect) experimental data.
 
Our calculated quantities, band gap and cohesion, are key inputs into more qualitative and approximate theories, such as tight-binding and k$\cdot$p, as well as into experimental interpretations. In particular, there is a clear path how our results can be modified to accommodate dielectric environments which enter the experiments, leaving the core of our QMC results unchanged. This can  be done by Bethe-Salpeter modeling or by use of model dielectric screening~\cite{veniard_17} to find the appropriate value of the exciton binding energy.
Hence, our explicitly many-body QMC results provide a reference ground for further studies on phosphorene based on strain and layer engineering as well as chemical doping and structural defects and indeed in any other 2D material.
We have also demonstrated that these cutting-edge calculations are now feasible for a range of 2D systems and we expect the QMC methods to find a place at the top of the list of the toolkit for studying 2D systems. This is underscored by the fact 
that the scatter of predicted values for electronic parameters 
is significant not only in phosphorene, but in other 2D materials as well.~\cite{cho_16,veniard_17,berkelbach_18} 

\begin{acknowledgments}
R.D., K.T., and I.S. were supported by APVV-15-0759, VEGA-2/0162/15 and VEGA-2/0123/18 projects. 
Contributions by L.M. have been supported by the NSF grant DMR-1410639 and by XSEDE computer time allocation at TACC. 
T.F. was supported by GRK  Grant  No.   1570,  and  the  International
Doctorate  Program  Topological  Insulators  of  the  Elite
Network  of  Bavaria.   J. F. acknowledges funding from 
the European Union's Horizon 2020 research and innovation program under grant agreement No.  696656, and 
DFG SFB 1277 (B07). 
Total estimated amount of computation time is in excess of 30 million core hours on current architecture.
T. F. and J. F acknowledge the Gauss Centre for
Supercomputing (www.gauss-centre.eu) for funding
this  project  by  providing  computing  time  on  the  GCS
Supercomputer  SuperMUC  at  Leibniz  Supercomputing
Centre (LRZ, www.lrz.de).
R.D., K.T., and I.S. acknowledge that the results of this research have been obtained using the DECI resources Beskow at the PDC Center for High Performance Computing at the KTH Royal Institute of Technology, Stockholm, Sweden from the PRACE project NaM2D and Sisu based in Finland at CSC IT Center for Science with support from the PRACE DoCSiNaP project.
We also gratefully acknowledge use of the Hitachi SR16000/M1 supercomputer system at CCMS/IMR, Tohoku University, Japan. We also acknowledge useful discussions with Martin Gmitra and Alexey Chernikov.
\end{acknowledgments}

\bibliography{paper}
\onecolumngrid
\newpage


\section*{Supplementary material}

# Supplementary Information to: Many-body quantum Monte Carlo study of 2D materials: cohesion and band gap in single-layer phosphorene

T. Frank,[1] R. Derian,[2] K. Tokár,[2] L. Mitas,[3] J. Fabian,[1, *] and I. Štich[2, 4, 5, †]

[1] *University Regensburg, Institute for Theoretical Physics, 93040 Regensburg, Germany*
[2] *Center for Computational Materials Science, Institute of Physics, Slovak Academy of Sciences, 84511 Bratislava, Slovakia*
[3] *Department of Physics, North Carolina State University, Raleigh, NC 27695-8202*
[4] *Institute of Informatics, Slovak Acad. of Sciences, 845 07 Bratislava, Slovakia*
[5] *Department of Natural Sciences, University of Ss. Cyril and Methodius, 917 01 Trnava, Slovakia.*

## I. SIMULATION DETAILS

The electronic fundamental band gaps $\Delta_f$, were determined for both, extended and cluster systems with B3LYP[1] DFT orbitals and also extended systems with PBE[2,3] DFT orbitals. Lattice parameters were fixed to the experimental values of black phosphorus crystal[4,5].

The fixed-node QMC[6,7] electronic gap calculations were performed in three steps:
1) the trial wave functions for ground- and excited-states were constructed from DFT B3LYP[1] and PBE[2,3] single-particle DFT wave functions. Note that, unlike the GGA-type PBE, the hybrid B3LYP functional yields a fairly realistic band gap of 1.7 eV already at the DFT level (see Table I of the main text). Note also that no charge transfer is associated with the HOMO→LUMO transition, indicating that range-separated hybrids, such as CAM-B3LYP functional[8], may not be needed.
2) the ground state trial wave function constructed from the DFT wave functions was optimized using VMC (variational Monte Carlo) techniques, and
3) finally fundamental gaps were computed from DMC (diffusion Monte Carlo) energies of ground- and excited-states as first singlet-singlet vertical excitation energies, where always the Jastrow factor of the ground state was used:

$$\Delta_f \approx E_v^s = E_1^s - E_0^s \quad ,$$

with $E_0$, $E_1$ being, respectively, the ground- and the first excited-states obtained by fixed-node DMC[6] (PBE/DMC and B3LYP/DMC) not allowing any relaxation of the DFT nodal hypersurfaces due to the HOMO→LUMO electron excitation.

Our QMC gap estimates use the vertical transition approximation, i.e. they neglect all the adiabatic, vibronic and zero-point vibrational energy effects. Such approximations tend to increase the gap value compared to the experiments, but on much smaller scale than that of importance here.

---

* jaroslav.fabian@ur.de
† ivan.stich@savba.sk

For DFT modeling we use the `CRYSTAL` code[9], while for all VMC and DMC calculations the `QWalk` code[10] was used. In all calculations the atomic cores were replaced by effective core potentials (ECP)[11] used in combination with a VTZ basis set.

Dynamical correlations were explicitly built in into the trial wave functions via isotropic Schmidt-Moskowitz Jastrow factors[7,12], including electron-electron and electron-nucleus correlations. Parameters of the trial wave functions[7] were optimized by minimizing linear combination of energy and variance[13]. The T-moves scheme[14] was used to keep the calculations with ECPs variational. Our systems are 2D+h in nature, i.e. while being 2D, they also have a finite thickness h. In order to take that structural characteristics into account, the Yeh-Berkowitz[15] modification of 3D Ewald summation technique for systems with a slab geometry that are periodic in two dimensions and have a finite length in the third dimension was adopted.

Two different types of finite-size scaling were performed:
1) a series of $L \times L$ supercell approximants where $L$ goes from 1 to 6, see Figure 1 of the main text, and
2) $4 \times 4$ and $5 \times 5$ H-terminated cluster approximants, see Figure 1 of the main text.
More details of the finite-size scaling are given in Sec. III.

The choice of these two types of models (periodic and finite) was driven by difficulties in estimating the band gap in the periodic setting. As explained and demonstrated below on solid data, the periodic calculations for excited states converge to the thermodynamic limit very slowly. Remarkably, we find that the cluster models converge to the infinite limit much faster and with much smaller variations of extrapolated estimators.

## II. COHESION ENERGY

We start the estimate of the cohesion energy from the $P_4$ molecule that by bonding patterns and energetics is very close to white and red phosphorus bulk crystals. $P_4$ is also the dominant sublimation product from phosphorus solids. Therefore we will use the following thermodynamic path molecule $P_4 \leftrightarrow$ black phosphorus bulk $\leftrightarrow$ phosphorene to find the cohesion. For $P_4$ binding we

initially use very accurate calculations and experimental data from the W4 testing set[16]. Coupled Cluster extrapolation to infinite basis and also experiment give an atomization energy of 285.03 kcal/mol (experiment being within 1 kcal/mol) and a zero point energy (ZPE) of 0.98 kcal/mol per atom. Our best Coupled Cluster extrapolated estimation is within 1.5 kcal/mol of the W4 value, see Table I. We also estimate the core-valence correlation and relativity corrections of −0.4 kcal/mol per atom[17] from calculations of $P_2$. Therefore the $P_4$ atomization energy (bottom of the well, infinitely heavy nuclei) at $T = 0$ K is 286.95 kcal/mol or 3.110(9) eV/atom.

Cohesion of the bulk black phosphorus can be estimated from the sublimation energy of black phosphorus $(P_4) \to$ gas$(P_4)$. We take a value of the heat of sublimation of 25.5(2.5) kJ/mol per atom or 0.26 eV/atom[18] (or almost the same value from an alternative experiment that estimates the cohesion energy difference between white and black bulks to be 21.2(2.5) kJ/mol and the heat of sublimation from white phosphorus to gas of 3.4 kcal/mol per $P_4$[19] so that we sum it to 25.5(2.5) kJ/mol = 0.26 eV/atom). Therefore the bulk black phosphorus cohesive energy with infinitely heavy nuclei is about 3.37(9) eV/atom (we assume that ZPE/atom is the same as in $P_4$ and this assumption also increases the error bar).

Bulk black phosphorus consists of stacked layers of phosphorene that are bounded by $\approx 0.08$ eV/atom[5]. Therefore a reasonable estimate for phosphorene cohesion that we can infer from this data is

$$E_{\text{coh},T\approx 0,\infty\text{nuclei}}^{\text{phosphorene}} = 3.29(9) \text{ eV/atom.}$$

We neglected relaxation of phosphorene layer in vacuum as compared to the layers in black phosphorus bulk but that is probably within the approximately estimated error bar.

Let us now analyze the fixed node diffusion Monte Carlo (FNDMC) calculations and results. Single FNDMC atom with PBE DFT orbitals, Burkatzki-Filippi-Dolg pseudopotentials[20], and single reference is

$$E_{\text{PBE FNDMC}}^{\text{single atom}} = -6.4740(4) \text{ Ha}$$

or

$$E_{\text{B3LYP FNDMC}}^{\text{single atom}} = -6.4737(4) \text{ Ha}$$

with B3LYP nodal hypersurfaces.

Results of the FNDMC phosphorene supercell calculations are shown in Supplementary Figure. 1 and the $N \to \infty$ extrapolated per atom energy is

$$E_{\text{PBE FNDMC}}^{\text{phosphorene}} = -6.59074(2) \text{ Ha} \to$$
$$E_{\text{coh PBE FNDMC}} \approx 3.178(2) \text{ eV/atom}$$

with PBE nodal hypersurfaces and

$$E_{\text{B3LYP FNDMC}}^{\text{phosphorene}} = -6.59109(2) \text{ Ha} \to$$
$$E_{\text{coh B3LYP FNDMC}} \approx 3.194(3) \text{ eV/atom}$$

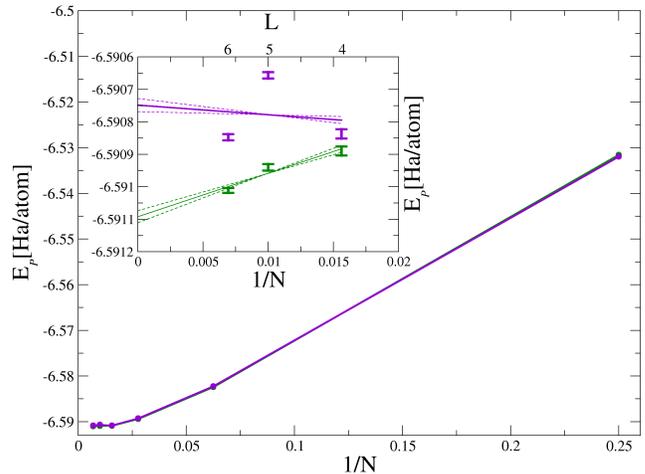

Supplementary Figure 1. Finite-size scaling of fixed-node DMC energy per atom in phosphorene in PBE/DMC (purple line) and B3LYP/DMC (green line) treatments. The error bars are smaller than the size of the points. The inset shows the zoom-in of the linear scaling for large $N$ with extrapolation to infinite system size.

Supplementary Table I. Basis set dependence and complete basis set (CBS) limit of CCSD(T) energies determined with Burkatzki-Filippi-Dolg pseudopotentials[20], in Hartree energies, of the P atom and the $P_4$ cluster.

| basis set | P atom | $P_4$ cluster |
|---|---|---|
| cc-pVTZ | −6.468765433 | −26.28345057 |
| cc-pVQZ | −6.475058531 | −26.34037340 |
| cc-pV5Z | −6.476654877 | −26.35692173 |
| CBS | −6.477197450 | −26.36370441 |

with B3LYP nodal hypersurfaces.

Both energies have fixed-node errors. We approximately estimate them from results of exact CCSD(T) in complete basis set limit (CBS) for the P atom and $P_4$ cluster since its cohesion/binding and bonding patterns are similar to phosphorene. The results of Burkatzki-Filippi-Dolg pseudopotential[20] calculations for a P atom and a $P_4$ cluster are in Supplementary Table I. To estimate the fixed-node error we need, in addition to he DMC atomic energies, also the $P_4$ energies, which we calculate as:

$$E_{\text{PBE FNDMC}}^{P_4} = -26.3376(2) \text{ Ha}$$

and

$$E_{\text{B3LYP FNDMC}}^{P_4} = -26.3365(2) \text{ Ha}$$

for PBE and B3LYP nodal hypersurfaces, respectively.

This gives the fixed-node errors of 0.095 eV/atom (0.088 eV/atom) and 0.185 eV/atom (0.178 eV/atom) in the atom and P$_4$ cluster with B3LYP (PBE) nodal hypersurfaces, respectively. Therefore we assume the FN differential error correction by about (0.178 − 0.088) eV/atom = 0.09 eV/atom at the PBE/DMC level and (0.185 − 0.095) eV/atom = 0.09 eV/atom at the B3LYP/DMC. Our corrected estimate of the cohesion energy is:

$$E_{\text{coh,PBE FNDMC,FN−corrected}} = 3.268(4) \text{ eV/atom},$$

and

$$E_{\text{coh,B3LYP FNDMC,FN−corrected}} = 3.284(6) \text{ eV/atom},$$

where the uncertainty is coming fully from the fixed-node correction since the statistical error bars are well below this. These fixed-node corrected cohesion energies compare also favorably with the estimate of 3.29(9) eV/atom given in the introduction.

## III. FINITE-SIZE SCALING

### A. HOMO-LUMO and $\Delta_f$ gap

The fundamental gap can be determined from total energies as[6]:

$$\Delta_f = (E_{N_e+1} − E_{N_e}) − (E_{N_e} − E_{N_e−1}) \quad (1)$$

with $N_e$ being the number of electrons. In the GGA approximation, $\Delta_f$ converges to the DFT HOMO-LUMO gap in the $N_e \to \infty$ limit and demonstrates that in the infinite limit at the DFT GGA level there is no interaction between the electron and the hole in the exciton and the HOMO-LUMO gap corresponds to the fundamental gap, $\Delta_f$. Such a conclusion is also valid in the generalized Kohn-Sham theory[21], i.e. also for hybrid functionals used in the main text. This is a key property for understanding our QMC cluster results in the main text. However, since the LUMO level is constructed identically for finite and infinite systems, the DMC cluster results will, in the infinite system size limit, converge to the true DMC value of $\Delta_f$. In addition, due to the confinement of the excitation, which can only increase the band gap, the cluster results will provide a strict upper limit for our extended system results.

### B. Finite size scaling of QMC gaps

An analytical formula for the asymptotic finite-size scaling of the band gap is not known. In the main text we use the scaling versus $1/N$. However, our calculated gap values could equally well be fitted with other fitting formulas, such as for instance with any $1/\sqrt[n]{N}$ formula. For example Drummond[22] used $1/\sqrt[2]{N}$. Indeed, fitting

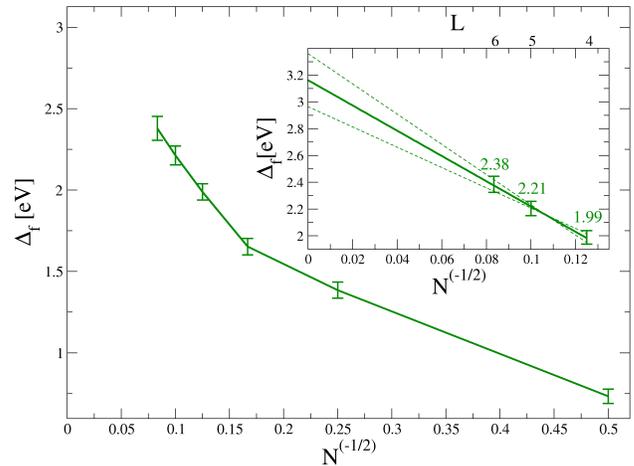

Supplementary Figure 2. Finite-size scaling of DMC/B3LYP fundamental gap, i.e. the gap values from Figure 2 of the main text, versus $1/\sqrt{N}$. Inset shows a zoom and extrapolation for the three largest supercells.

our data versus $1/\sqrt[2]{N}$, Supplementary Figure 2, yields a fit that extrapolates to $\Delta_f$ of 3.08 eV. This, though, is in disagreement with the upper bound provided by the cluster calculations which exclude any higher root extrapolation than 1, i.e., the linear finite-size extrapolation versus $1/N$ is used the main text. Further analysis of the behavior for larger sizes supports these conclusions and provides therefore a rather solid estimation. This is further supported by the analysis of the cohesion energy that shows that the ground state is converged at sizes 4×4 or larger (see main text Fig. 3). The convergence of the excited state in periodic setting is slower due to the presence of vacuum in the direction orthogonal to the phosphorene slab. This enables uninhibited restructuring of the charge in this direction and thus generates long-range (artificial) interactions. In particular, the excited state, as it is clear from Figure 2 of the main text, shows basically an antibonding pattern along one of the P–P bonds that is almost orthogonal to the phosphorene plane ($xy$-direction) and therefore possesses a significant tail in the $z$−direction. The excited state is then a repetition of the same bonding pattern at all symmetry-related bonds since the excitation corresponds to the Γ-point symmetry. This results in arrays of dipoles on both sides of the slab and contributes significantly to the potential energy. Indeed, this is amply visible in the corresponding HF energies for the excited states (see Sec. C). The convergence to the thermodynamic limit is therefore very slow and extrapolation is quite difficult and very demanding on computational resources.

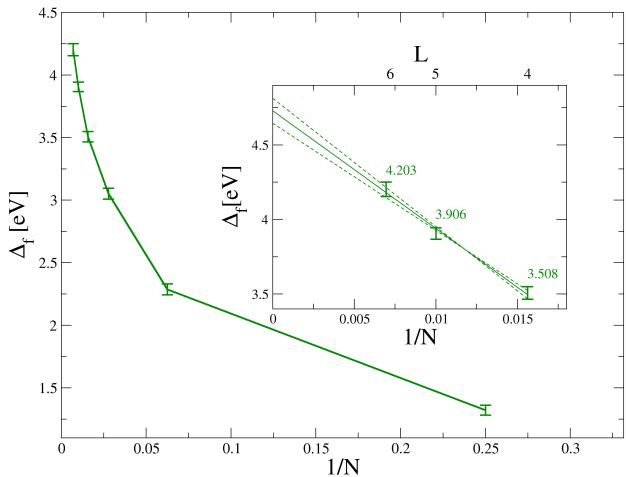

Supplementary Figure 3. Finite-size scaling of the Hartree-Fock fundamental gap versus $1/N$. Inset shows a zoom and extrapolation for the three largest supercells.

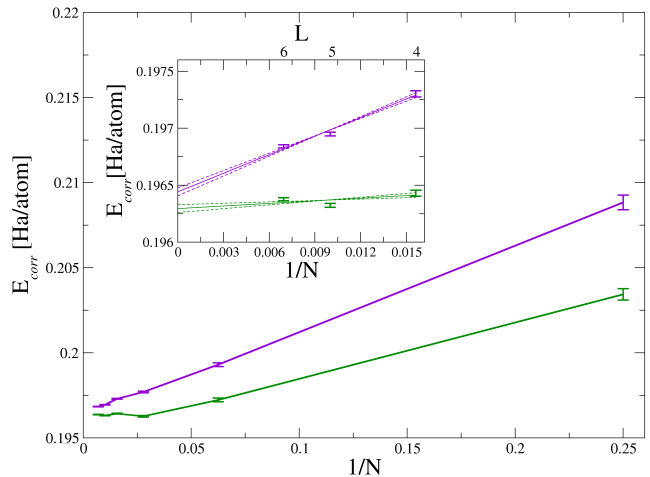

Supplementary Figure 4. Finite-size scaling of the correlation energies of the ground-state $S_0$ (green line) and the first excited-state $S_1$ (purple line) versus $1/N$.

### C. Hartree-Fock band gaps

The main contribution to the band gap, the Hartree-Fock part, is plotted in Supplementary Figure 3. This was calculated from the VMC wave function with B3LYP orbitals, setting the Jastrow correlation factor to 1. The Hartree-Fock band gap is extrapolated to 4.720(80) eV in the infinite size limit.

### D. Finite-size scaling of correlation energies

Finite-size scaling of correlation energies of both ground- and excited-states are shown in Supplementary Figure 4. The correlation energy of the ground-state are essentially converged at the $4 \times 4$ supercell size while that of the excited state appears to converge more slowly but also appears to show sign of being converged for the $6 \times 6$ supercell size. The finite-size extrapolated values of correlation energies are 0.19629(3) Ha/atom and 0.19644(3) Ha/atom for the ground- and excited state, respectively. Since the correlation energies converge faster than the Hartree-Fock energies, see Supplementary figure 3, larger approximants could be constructed simply by adding the correlation energies to the Hartree-Fock energies.

## IV. EXTRAPOLATIONS BASED ON CALCULATED CORRELATION ENERGIES

Having shown in Sec. III D that the correlation energies tend to saturate at smaller system sizes than the Hartree-Fock energies, we now show that this fact may be used to estimate energies and band gaps for larger approximants. We applied this idea to the $6 \times 6$ approximant in the cluster B3LYP/DMC treatment. From the smaller $4 \times 4$ and $5 \times 5$ DMC cluster calculations the (converged) correlation energies per P and H atom can be estimated, provided the Hartree-Fock energies are also known, Supplementary Table II. The relevant numbers are compiled in Supplementary Table II. From the energies compiled in the Supplementary Table II we estimate a fundamental gap of $\Delta_f^{6\times6,\ \text{clst, est}} = 2.580(150)$ eV, which is almost within the error bar of the extrapolated value. It confirms that the band gaps of the clusters beyond $4 \times 4$ are much more weakly dependent on the size. This enables us to cross check the values from periodic calculations that show much stronger variation with the system size.

Consistency of cluster and periodic calculations can be ascertained also for the total energies per atom and the subsequent cohesion estimation. For clusters $L \times L$ we can use the following total energy model $E_L = N_P E_P + N_H E_H$ that adds energies of corresponding numbers of P and H atoms (see Table II). Based on this we can find linear relationships between different $E_L$ such as, for example, $E_6 = (12/5)E_5 - (3/2)E_4$, etc. Influence of finite sizes is as we mentioned much smaller for clusters than for the periodic calculations. The data enables us therefore to estimate the corresponding total energy per P atom in the thermodynamic limit also in the cluster setting giving the following value

$$E_{\text{FNDMC B3LYP}}^{\text{phosphorene, cluster}} = \frac{1}{80}(4E_5 - 5E_4) = -6.5904(1) \text{ Ha}$$

It compares very favorably with the result from our periodic calculations (compare Sec. II, difference $\approx$ 0.02 eV/atom) and shows thus remarkably close agreement between these estimations despite significant difference in boundary conditions of the two models.

Supplementary Table II. Total and correlation energies of the ground- (GS) and excited-state (ES) for clusters, in Ha. The Hartree-Fock energies were calculated from Slater determinants with B3LYP orbitals via VMC, setting the Jastrow factor to 1. Bold numbers are estimated values based on computed energies of smaller, 4 × 4 and 5 × 5 clusters.

| $L$ | # atoms | | total energy | | HF energy | | correlation energy | |
|---|---|---|---|---|---|---|---|---|
| | P | H | GS | ES | GS | ES | GS | ES |
| 4 | 64 | 24 | −435.9029(9) | −435.8046(8) | −423.2858(7) | −423.1230(7) | 12.6171(16) | 12.6816(15) |
| 5 | 100 | 30 | −676.6902(10) | −676.5955(9) | −657.0435(10) | −656.8789(10) | 19.6467(20) | 19.7166(19) |
| 6 | 144 | 36 | **−970.1994(40)** | **−970.1043(39)** | −941.9730(13) | −941.8069(13) | **28.2264(37)** | **28.2974(36)** |

Supplementary Table III. Tabulated values of calculated band gaps in eV. The Hartree-Fock values correspond to band gaps calculated from B3LYP wave functions via VMC setting the Jastrow factor to 1.

| $L$ | 1 | 2 | 3 | 4 | 5 | 6 | ∞ |
|---|---|---|---|---|---|---|---|
| HF@B3LYP-periodic | 1.320(40) | 2.390(44) | 3.140(44) | 3.510(42) | 3.890(38) | 4.250(48) | 4.720(80) |
| HF@B3LYP-cluster | - | - | - | 4.430(27) | 4.480(38) | 4.520(52) | 4.580(80) |
| DMC@B3LYP-periodic | 0.730(44) | 1.390(49) | 1.650(50) | 1.990(50) | 2.210(53) | 2.380(60) | 2.680(100) |
| DMC@PBE-periodic | 0.740(48) | 1.370(49) | 1.660(52) | 1.760(50) | 1.960(53) | 2.220(71) | 2.540(120) |
| DMC@B3LYP-cluster | - | - | - | 2.670(47) | 2.580(52) | - | 2.410(170) |

## V. TABULATED DMC GAPS

All computed gap values are compiled in Supplementary Table III.


\pagestyle{empty}
   
\end{document}